# High electron density $\beta$-(Al$_{0.18}$Ga$_{0.82}$)$_2$O$_3$/Ga$_2$O$_3$ modulation doping using ultra-thin (1 nm) spacer layer

Nidhin Kurian Kalarickal[1], Zhanbo Xia[1], Joe Mcglone[1], Yumo Liu[2], Wyatt Moore[1], Aaron Arehart[1], Steve Ringel[1, 2], Siddharth Rajan[1, 2]

[1]*Department of Electrical and Computer Engineering, The Ohio State University, Columbus, OH 43210, USA*

[2] *Department of Materials Science and Engineering, The Ohio State University, Columbus, OH 43210, USA*

We report on the design and demonstration of $\beta$-(AlGa)$_2$O$_3$/Ga$_2$O$_3$ modulation doped heterostructures to achieve high sheet charge density. The use of a thin spacer layer between the Si delta-doping and heterojunction interface was investigated in $\beta$-(AlGa)$_2$O$_3$/Ga$_2$O$_3$ modulation doped structures. We find that that this strategy enables higher 2DEG sheet charge density up to $6.1 \times 10^{12}$ cm$^{-2}$ with mobility of 147 cm$^2$/Vs. The presence of a degenerate 2DEG channel was confirmed by the measurement of low temperature effective mobility of 378 cm$^2$/V-s and a lack of carrier freeze out from low temperature capacitance voltage measurements. The electron density of $6.1 \times 10^{12}$ cm$^{-2}$ is the highest reported sheet charge density obtained without parallel conduction channels in an (AlGa)$_2$O$_3$/ Ga$_2$O$_3$ heterostructure system.

With a high theoretical breakdown field strength of 8 MV/cm [1,2], $\beta$-Ga$_2$O$_3$ has the potential to be useful in several high frequency[3,4] and power switching applications[5,6]. The high break down field enables shrinking the overall device footprint which results in improved frequency performance for power switching devices and increased output power density for RF power amplifiers. Besides the superior breakdown field strength, the availability of native $\beta$-Ga$_2$O$_3$ substrates [7–9] enables high quality epitaxial growth using techniques such as molecular beam epitaxy [10,11], metal organic chemical vapor deposition [12,13], halide vapor phase epitaxy [14,15] and pulsed laser deposition [16,17].

For lateral power devices it is essential for the channel to be placed close to the gate. Firstly, for enhancement mode devices (which are preferred for power electronics), a lower gate-to-channel spacing leads to higher gate-to-channel capacitance, and therefore enables higher sheet charge density for the same gate voltage swing. In addition, a scaled channel allows for better control of gate-drain electric field and lateral scaling of gate length. The former is important in achieving high average breakdown field strength while the latter is crucial in improving the frequency of operation in RF power amplifiers and reducing on resistance in power switching devices.

Lateral devices like $\beta$-Ga$_2$O$_3$ MESFETs for high frequency application have been demonstrated with high on/off ratio and breakdown voltage but the performance of these devices is mainly limited by the low mobility (50-90 cm$^2$/V-s) [3,4]. Introducing 2-dimensional electron gas (2DEG) channels are the optimal solution since they provide the highest sheet charge density for a given gate voltage swing, while also providing superior transport properties. $\beta$-(AlGa)$_2$O$_3$/Ga$_2$O$_3$ modulation doped field effect transistors (MODFETs) are attractive in this respect since they can enable a 2DEG with excellent transport properties. $\beta$-(AlGa)$_2$O$_3$/Ga$_2$O$_3$ MODFETs with mobility as high as 180 cm$^2$/V-s have been demonstrated recently [18], but these devices are mainly limited by a low conduction band offset (0.3- 0.4 eV) resulting in low sheet charge density (~2x10$^{12}$ cm$^{-2}$). Since $\beta$-(AlGa)$_2$O$_3$/Ga$_2$O$_3$ heterostructure is predicated to have a type I band alignment [19], the conduction band offset is determined by the Al mole fraction. Density functional theory based calculations show that Al composition in excess of 50% is required to achieve a band offset of 1 eV [16,19]. Achieving such high Al composition is limited by phase separation of (AlGa)$_2$O$_3$ above 30% of Al composition due to the dissimilar parent crystal structures [20]. Currently, high quality $\beta$-(AlGa)$_2$O$_3$ films are limited to below 30% of Al incorporation. This fixes the achievable conduction band offset in $\beta$-(AlGa)$_2$O$_3$/Ga$_2$O$_3$ system to around 0.4 eV. In a delta doped modulation doped heterostructure the maximum permissible 2DEG charge density ($n_s^{max}$) without the formation of a parallel channel is given by

$$n_s^{max} = \frac{\epsilon \Delta E_c}{q(d_s + \Delta d)},$$

where $\Delta E_c$ is the conduction band offset between $\beta$-Ga$_2$O$_3$ and $\beta$-(AlGa)$_2$O$_3$, $d_s$ is the spacer thickness (Figure.1 (a)), $\Delta d$ is the effective quantum capacitance distance $\left(\frac{\epsilon \pi \hbar^2}{m^* q^2}\right)$ and $\epsilon$ is the dielectric permittivity. For a fixed value of $\Delta E_c$ the only way to increase the charge density is by reducing the spacer thickness $d_s$. Increasing the charge density is crucial in reducing the on resistance and increasing drain current in MODFET devices. Additionally, an increase in charge density would result in improved screening of the strong polar optical phonon scattering increasing the mobility [21]. Figure.1 (b) shows the simulated band diagram of a $\delta$ doped ($9.5 \times 10^{12}$ cm$^{-2}$) $\beta$-(AlGa)$_2$O$_3$/Ga$_2$O$_3$ MODFET with a thin spacer layer (1 nm) showing the formation of a 2DEG ($6.1 \times 10^{12}$ cm$^{-2}$) without the presence of parallel conducting channels.

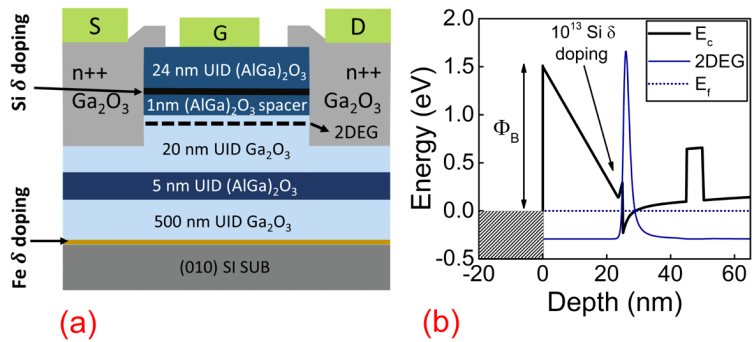

Figure.1 (a) Epitaxial diagram of $\beta$-(AlGa)$_2$O$_3$/Ga$_2$O$_3$ thin spacer modfet, (b) band diagram along vertical cutline through the gate.

In this report we demonstrate a method to increase sheet charge density in $\beta$-(AlGa)$_2$O$_3$/Ga$_2$O$_3$ modfets by aggressively scaling the spacer layer thickness to 1 nm. In the case of $\beta$-Ga$_2$O$_3$ since the polar optical phonon scattering limited mobility is already low (200 cm$^2$/V-s [22]), the increased remote ionized impurity scattering could be expected to have less impact on the total mobility. Therefore using a thin spacer could enable higher 2DEG sheet charge density without significant impact on 2DEG mobility. These ideas are confirmed by scattering rate calculations and experimental results discussed later in this letter (Figure.4 (a)).

**Epitaxial growth:** The $\beta$-(AlGa)$_2$O$_3$/Ga$_2$O$_3$ modulation doped heterostructure was epitaxially grown in a Riber M7 MBE system equipped with a Veeco oxygen plasma source. The growth utilized Ga limited conditions with a Ga beam equivalent pressure (BEP) of $8 \times 10^{-8}$ Torr, O$_2$ pressure of $1.5 \times 10^{-5}$ Torr, RF plasma power of 300 W and a substrate temperature of 610 °C (pyrometer calibrated to Si emissivity). We grew the epitaxial structure shown in Figure.1 (a) on an Fe-doped $\beta$-Ga$_2$O$_3$ (010) substrate (Novel crystal Technology [23]). Fe delta-doping at the growth interface was used to eliminate interface impurity related parasitic conduction paths. Si delta doping was carried out by opening the Si shutter for 2.5 s at a cell temperature of 950 °C [24]. Al composition of 17% was estimated based on peak separation from HRXRD (Figure.2 (a)) and an RMS roughness of 0.6 nm was observed from AFM (not shown). The growth temperature used here, 610 °C, was chosen to be lower than our typical optimal growth temperatures to reduce the spread of the Si delta sheet by diffusion through the thin spacer layer.

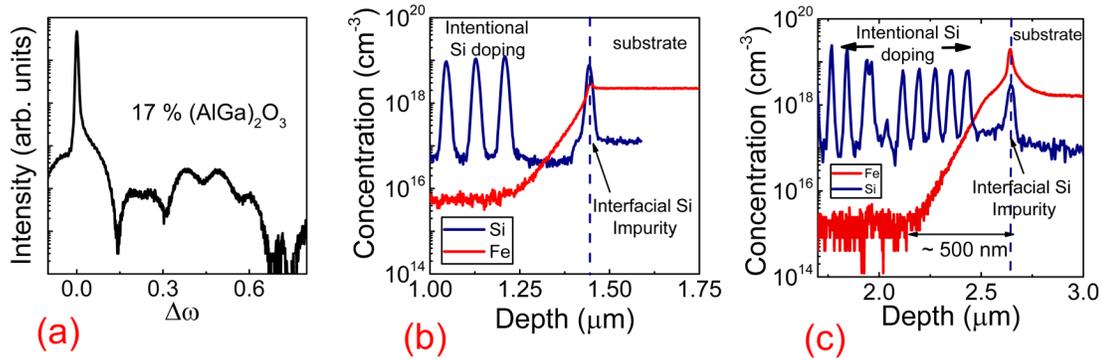

Figure.2 (a) High resolution XRD showing 18% (AlGa)$_2$O$_3$, (b) SIMS profile showing the formation of Si impurity channel at the growth interface, (c) SIMS profile showing compenstation of Si impurity channel by introducing intentional Fe $\delta$ doping.

We added an intentional Fe $\delta$ doping layer to compensate the parasitic channel formed at the growth interface due to Si contamination. Figure. 2 (b) shows the secondary ion mass spectrometry (SIMS, Evans Analytical Group [25]) profile of a sample without the Fe $\delta$ doped layer showing the formation of a Si impurity channel. This secondary channel prevents pinch off in transistor devices and leads to significant buffer leakage. Introducing acceptor Fe atoms at the growth interface leads to compensation and removal of the parasitic channel. Fe $\delta$ doping was achieved by opening the Fe shutter for 8s (1140 °C) at the start of the growth and immediately cooling down the cell (Figure.2 (c)). The epitaxial surface riding tendency of Fe in $\beta$-Ga$_2$O$_3$ leads to a significant Fe tail of around 500 nm [26]. Therefore, a minimum buffer thickness of 500 nm is required so that the active channel is spatially separated from the Fe tail. A 5 nm (AlGa)$_2$O$_3$ back barrier was added in addition to the buffer layer to further improve this spatial separation (Figure.1 (a)).

**Charge and transport:** We fabricated ohmic contacts on the $\beta$-(AlGa)$_2$O$_3$/Ga$_2$O$_3$ heterostructure by etching away the contact regions and growing heavily doped n-type Ga$_2$O$_3$. Ti/Au (30 nm/ 100 nm) ohmic metal contact was later deposited using e-beam evaporation and annealed at 470 °C in N$_2$ ambient. We carried out device isolation using a mesa structure etched using BCl$_3$/Ar based dry etch (BCl$_3$/Ar flow of 20/5 sccm, ICP/RIE power of 200/30 W and process pressure of 15 mTorr). Gate contacts were patterned using stepper lithography followed by e-beam evaporation of Ni/Au (30 nm/ 100 nm) schottky metal stack.

We estimated an ohmic contact resistance of 3.2 $\Omega$.mm and sheet resistance of 8.3 $K\Omega$/■ using transfer length measurements. These match fairly well with hall measurements carried out on van der pauw structures, from which we extracted a Hall mobility of 147 cm$^2$/V-s and a sheet charge density of $4.4 \times 10^{12}$ cm$^{-2}$. The Hall mobility for these structures is comparable to MODFETs with thick spacer layers ($d_s > 4$ nm) (Figure.3 (d)). Capacitance-voltage measurement (C-V) was utilized to probe the 2DEG channel both at room temperature and liquid nitrogen temperature (77 K) (Figure.3 (a, b)). Based on the C-V measurements we estimate a total sheet charge density of $6.1 \times 10^{12}$ cm$^{-2}$ at room temperature, and $5.8 \times 10^{12}$ cm$^{-2}$ at 77 K. Absence of significant carrier freeze out at cryogenic temperature suggests the presence of a degenerate 2DEG without parallel conduction in the (AlGa)$_2$O$_3$ layer. Previous reports showed that when parallel conducting channels were present, the spatial overlap of donors and carriers resulted in carrier freeze out at low temperature [27].

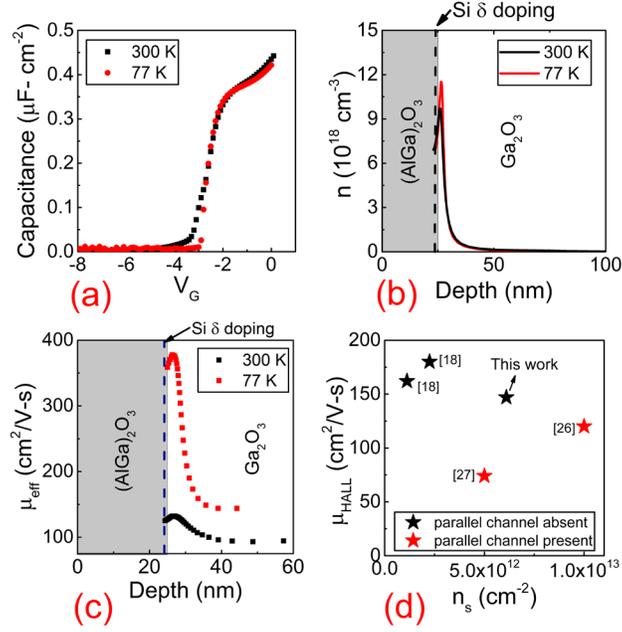

Figure.3 (a) Capacitance-voltage characteristics (100 KHz) measured at 300 K and 77 K, (b) extracted charge profile as a function of depth, (c) effective mobility as a function of depth, (d) benchmark plot comparing this result with previous reports of modulation doped $\beta$-(AlGa)$_2$O$_3$/Ga$_2$O$_3$ heterostructures [18,27,28].

We used large gate length structures (100 $\mu$m gate length) to measure the effective mobility as a function of sheet charge. The large gate length reduces the effect of source and drain access regions (2 $\mu$m each), so we can use the ideal gradual channel approximation to estimate the effective drift mobility as

$$\mu_{eff} = \frac{g_d(V_g) \times L_g}{q n_s(V_g) \times w}$$

where $g_d(V_g)$ is the drain conductance, $n_s(V_g)$ is the channel sheet charge density, $L_g$ is the gate length and $w$ is the gate width. Figure.3 (c) compares the measured effective mobility as a function of depth $\left(\frac{\epsilon}{C_{gs}}\right)$ at room temperature and 77 K. At room temperature, the effective mobility shows a peak value of 132 cm$^2$/V-s (4.2 $\times$ 10$^{12}$ cm$^{-2}$ sheet density) and 120 cm$^2$/V-s at zero gate bias (6.1 $\times$ 10$^{12}$ cm$^{-2}$ sheet density). As the temperature is lowered to 77 K, the mobility increases, with a peak mobility of 378 cm$^2$/V-s and a zero-bias mobility of 360 cm$^2$/V-s. The increased low temperature mobility, coupled with the absence of carrier freeze out confirms a degenerate channel with a mobility that is not limited by ionized impurity scattering (Figure.4 (a)). When compared with the lowest sheet resistance obtained in $\beta$-(AlGa)$_2$O$_3$/Ga$_2$O$_3$ MODFETs before this we find that we have a significantly lower room temperature sheet resistance due to the increased 2DEG density (Figure.3 (d)).

**Mobility modelling:** To understand the scattering mechanism that limits transport in the thin spacer MODFETs, we compared the experimental effective mobility with theoretical estimates. Scattering mechanisms considered include remote ionized impurity scattering (RI), polar optical phonon scattering (POP), background impurity scattering (BI) and interface roughness scattering (IR), which are expected to be the main limiting mechanisms for mobility. RI, BI and IR scattering limited mobility ($\mu_{RI}$, $\mu_{BI}$, $\mu_{IR}$) values were calculated based on the analysis in Ref [29] and [30]. We used the polar analysis in [31] to evaluate the optical phonon limited mobility ($\mu_{POP}$) and used Matheissen's rule to approximate the total scattering rate as a sum of the different mechanisms, giving

$$\frac{1}{\mu_{TOT}} = \frac{1}{\mu_{RI}} + \frac{1}{\mu_{POP}} + \frac{1}{\mu_{BI}} + \frac{1}{\mu_{IR}}$$

Table I gives a list of relevant material parameters used for mobility modelling. As shown in Figure.4 (a), the measured effective mobility values are in reasonable agreement with the calculated mobility over a range of sheet carrier density values. Coulombic scattering mechanisms like RI and BI which limits the low temperature mobility (Figure.4 (a)) accounts for only a fraction of the room temperature mobility. The average scattering time at room temperature is therefore dominated by polar optical phonon scattering even at a spacer thickness of 1 nm.

TABLE I Material parameters of $\beta$-Ga$_2$O$_3$

| Parameter | symbol | Value used |
|---|---|---|
| Effective mass | $m^*$ | 0.3 m$_o$ |
| DC dielectric constant | $\epsilon$ | $8.85 \times 10^{-13}$ F/cm |
| Optical phonon energy | $E_{op}$ | 50 meV |
| High frequency dielectric constant | $\epsilon_\infty$ | 3.6 |
| Donor sheet charge density | $n_{imp}^{2D}$ | $10^{13}$ cm$^{-2}$ |
| Background impurity concentration | $N_{back}^{3D}$ | $3.5 \times 10^{18}$ cm$^{-3}$ |
| Root mean square roughness | $\Delta$ | 1.2 nm |
| Correlation length | $L$ | 4 nm |

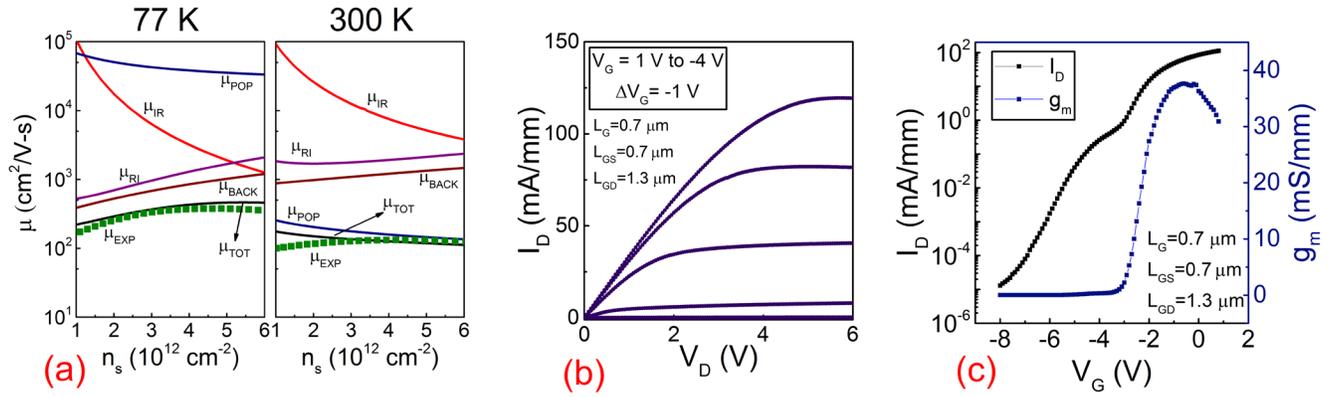

Figure.4 (a) Comparison of experimental effective mobility values with scattering theory at 77 K and 300 K, (b) Output characteristics of $\beta$-$(AlGa)_2O_3/Ga_2O_3$ thin spacer modfet, (c) Transfer characteristics of $\beta$-$(AlGa)_2O_3/Ga_2O_3$ thin spacer modfet measured at $V_D=5$ V.

**Three terminal devices:** Figure. 4 (b), (c) shows the output and transfer characteristics of a three terminal device with a gate length ($L_g$) of 0.7 $\mu m$, gate to source spacing ($L_{gs}$) of 0.7 $\mu$m and gate to drain spacing ($L_{gd}$) of 1.3 $\mu m$. The output characteristics show a peak drain current of 120 mA/mm which is the highest reported for a single channel $\beta$-$(AlGa)_2O_3/Ga_2O_3$ MODFET. Improvement in output characteristics is a direct result of the improvement in 2DEG sheet charge density. The device also shows a peak transconductance ($g_m$) of 38 mS/mm and a pinch off voltage of -8 V (Figure.4 (c)). The presence of Si impurity channel at the growth interface is visible as a second peak in the transfer characteristics at -4 V (Figure.4 (c)). We believe this may be because intentional Fe delta doping at the growth interface has not completely compensated interfacial donor impurities. Therefore even though the 2DEG channel is depleted at a gate bias of -3.5 V, the parasitic channel leads to a more negative pinch-off voltage (-8 V).

In conclusion we have showed that usage of a thin (1 nm) spacer layer is an effective strategy in $\beta$-$(AlGa)_2O_3/Ga_2O_3$ modulation doped heterstuctures. Hall mobility of 147 cm$^2$/V-s and 2DEG sheet charge density of $6.1 \times 10^{12}$ cm$^{-2}$ were demonstrated for Si $\delta$ doped thin spacer $\beta$-$(AlGa)_2O_3/Ga_2O_3$ MODFETs. Mobility modelling shows that even at 1 nm spacer thickness the 2DEG mobility is limited by polar optical phonons and not by remote ionized impurity scattering. Three terminal devices on thin spacer $\beta$-$(AlGa)_2O_3/Ga_2O_3$ MODFETs show a peak drain current of 120 mA/mm and a peak transconductance of 38 mS/mm. Future epitaxial structures using such thin spacer layers, or multiple channels formed by stacking a few such 2-dimensional electron gases could be attractive for power and high frequency transistors based on $\beta$-$Ga_2O_3$.

We acknowledge funding from the AFOSR under Grant no. FA9550-18-1-0479 (GAME MURI, program manager Dr. Ali Sayir) and from Northrop Grumman NG NEXT Basic Research (Dr.Vincent Gambin).